\documentclass[preprint,amsmath,amssymb,showkeys]{revtex4-1}
\usepackage{graphicx}
\renewcommand{\d}{{\rm d}}
\newcommand{\e}{{\rm e}}
\newcommand{\eref}[1]{(\ref{#1})}
\usepackage{graphicx}

\begin{document}

\title{Exact Solutions and Flow--Density Relations for %
a Cellular Automaton Variant of the Optimal Velocity Model %
with the Slow-to-Start Effect}
\altaffiliation{Accepted for publication in J.~Phys.~Soc.~Jpn.}

\author{Hideaki Ujino}
\email{ujino@nat.gunma-ct.ac.jp}
\affiliation{Gunma National College of Technology, %
580 Toriba, Maebashi, Gunma 371--8530, Japan}

\author{Tetsu Yajima}
\email{yajimat@is.utsunomiya-u.ac.jp}
\affiliation{Department of Information Science, 
Graduate School of Engineering, Utsunomiya University,
7--1--2 Yoto, Utsunomiya, Tochigi 321--8585, Japan}

\begin{abstract}
A set of exact solutions for a cellular automaton, which is a
hybrid of the optimal velocity and the slow-to-start 
models, is presented. The solutions allow coexistence of free flows
and jamming or slow clusters, which is observed in asymptotic behaviors
of numerically obtained spatio-temporal patterns.
An exact expression of the flow--density relation given by 
the exact solutions of the model agrees with an empirical formula
for numerically obtained flow--density relations.
\end{abstract}


\maketitle

\section{Introduction}

Studies on microscopic models for vehicle traffic
provided a good point of view on the phase transition
from the free traffic flow to the congested one. Related
self-driven many-particle systems have attracted
considerable interests not only from engineers but
also from physicists~\cite{Chowdhury2000,Helbing2001}.
Among such models, the optimal velocity (OV) model~\cite{Bando1995},
which is a car-following model describing an adaptation to 
the optimal velocity that depends on the headway
between two neighboring vehicles, is well-known for 
its successful rationalization of ``phantom traffic jams''
in the high-density regime.

Whereas the OV model consists of ordinary differential equations (ODE),
cellular automata (CA) such as the Nagel--Schreckenberg model~\cite{Nagel1992},
the elementary CA of Rule 184 (ECA184)~\cite{Wolfram1986}, the Fukui--Ishibashi
(FI) model~\cite{Fukui1996} and the slow-to-start (s2s)
model~\cite{Takayasu1993} are extensively used in analyses of traffic flow.
A way toward amalgamation of the ODE-type and the CA-type models was opened
by the discovery of the discrete OV (dOV) 
model~\cite{Takahashi2009} that provides 
an ultradiscretization~\cite{Tokihiro1996} of the OV model.
The resultant ultradiscrete OV (uOV) model includes both the ECA184 
and the FI model as its special cases. 
We should note here that 
another ultradiscretization~\cite{Kanai2009} of the OV model 
is developed out of the
ultradiscretization of the mKdV equation~\cite{Takahashi1997},
which also has an application to the traffic flow~\cite{Emmerich1998}.

Inspiration brought by 
the ultradiscretization of the OV model in ref.~\cite{Takahashi2009}
leads us to 
an ultradiscretizable hybrid of the OV and the s2s models,
which was named the optimal velocity model with
the slow-to-start effect, or shortly, the s2s--OV model~\cite{Oguma2009},
\begin{equation}
  \dfrac{\d x_k(t)}{\d t} 
  = v_0\Bigl(1+\dfrac{1}{t_0}\int_0^{t_0}
  \e^{-(\Delta x_k(t-t^\prime)-x_0)/\delta x}\d t^\prime
  \Bigr)^{-1}
  -v_0\bigl(1+\e^{x_0/\delta x}\bigr)^{-1},
  \label{eq:s2s--OV}
\end{equation}
where $t_0$ and $x_k(t)$ are the monitoring period and the position of the $k$-th car at the time $t$, respectively. 
The interval between the cars $k$ and $k+1$ is
denoted by $\Delta x_k(t)=x_{k+1}(t)-x_k(t)$.
Sensitivity to the interval is controlled by $\delta x>0$ and
$x_0$ means the length of the road occupied by a vehicle.
In the limit $t_0\rightarrow 0$, 
the s2s--OV model reduces to 
the Newell model~\cite{Newell1961}, 
which is a car-following model describing an retarded adaptation 
to the optimal velocity determined by the headway in the past.
The ultradiscrete limit of the s2s--OV model includes
a traffic-flow model of the CA-type,
\begin{equation}
  x_k^{n+1}=x_k^n+\min\Bigl(\min_{n^\prime=0}^{n_0}\bigl(\widetilde{\Delta}
  x_k^{n-n^\prime}\bigr),v_0\delta t\Bigr),
  \label{eq:s2s--OVCA}
\end{equation}
where the position of the $k$-th car $x_k^n$, $k=1,2,\cdots, K$, and 
the headway between the cars $k$ and $k+1$
at the $n$-th discrete time
$\widetilde{\Delta}x_k^{n}:=x_{k+1}^{n}-x_{k}^n-x_0\geq 0$
are integers. We should note that the headway $\widetilde{\Delta}x_k^{n}$
and the interval $\Delta x_k^{n}:=x_{k+1}^{n}-x_{k}^n=\widetilde{\Delta}x_k^{n}+x_0$
between the cars $k$ and $k+1$ are used in different meanings in this paper.
If all the initial headways $\widetilde{\Delta}x_k^{0}$ 
are non-negative, then car crash is prohibited at any time.
Positive integers $v_0$, $x_0$ and $\delta t$ means 
the speed limit, the size of a cell and the discrete time step, 
respectively. 
When we consider the time-evolution of the vehicles in the following,
the latter two
parameters, $x_0$ and $\delta t$, are fixed at one. 
Then eq.~\eref{eq:s2s--OVCA} reduces to
\begin{equation}
  x_k^{n+1}=x_k^n+\min\Bigl(\min_{n^\prime=0}^{n_0}\bigl(
  x_{k+1}^{n-n^\prime}-x_{k}^{n-n^\prime}-1\bigr),v_0\Bigr).
  \label{eq:s2s--OVCA2}
\end{equation}
We call a positive integer $n_0$ the monitoring period since the minimum
of the headway 
$\widetilde{\Delta}x_k^{n-n^\prime}:=x_{k+1}^{n-n^\prime}-x_k^{n-n^\prime}-x_0$
of $n_0+1$ discrete times, $n-n^\prime=n-n_0, n-n_0+1,\cdots, n$, 
is involved in the time evolution~\eref{eq:s2s--OVCA}. This CA describes
many cars running on a single lane highway in one direction, which is driven
by cautious drivers requiring enough headway to go on at least for
$n_0$ time steps before they accelerate their cars. 
Since the above CA~\eref{eq:s2s--OVCA} is an extension of the CA
case of the uOV model~\cite{Takahashi2009} incorporating the
s2s effect in the sense of delay in drivers' response, 
we call the CA~\eref{eq:s2s--OVCA} the CA variant of the
OV model with the s2s effect,  or the s2s--OVCA for short.

\begin{figure}[h]
\begin{center}
\includegraphics[width=60mm]{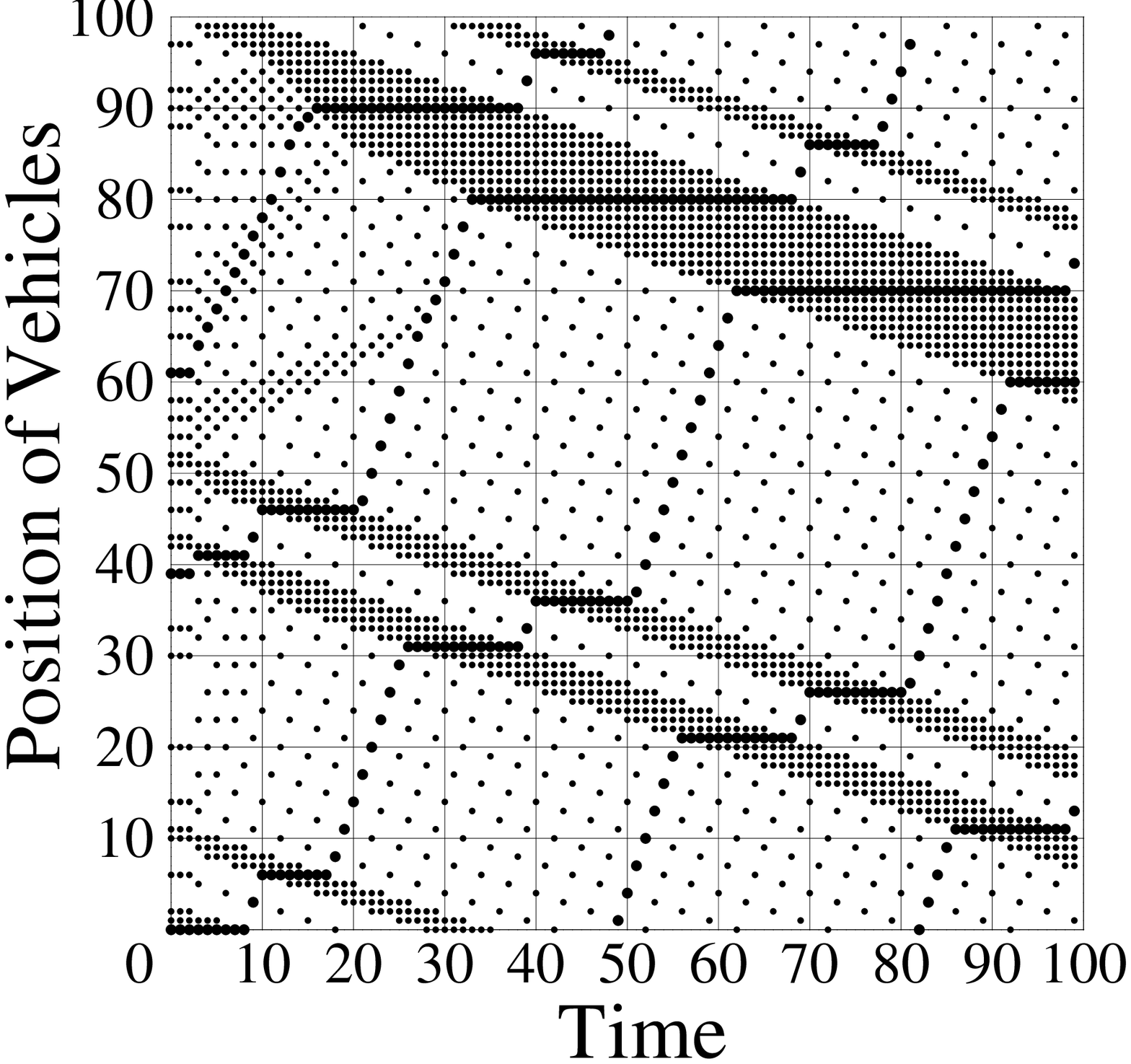}
\includegraphics[width=60mm]{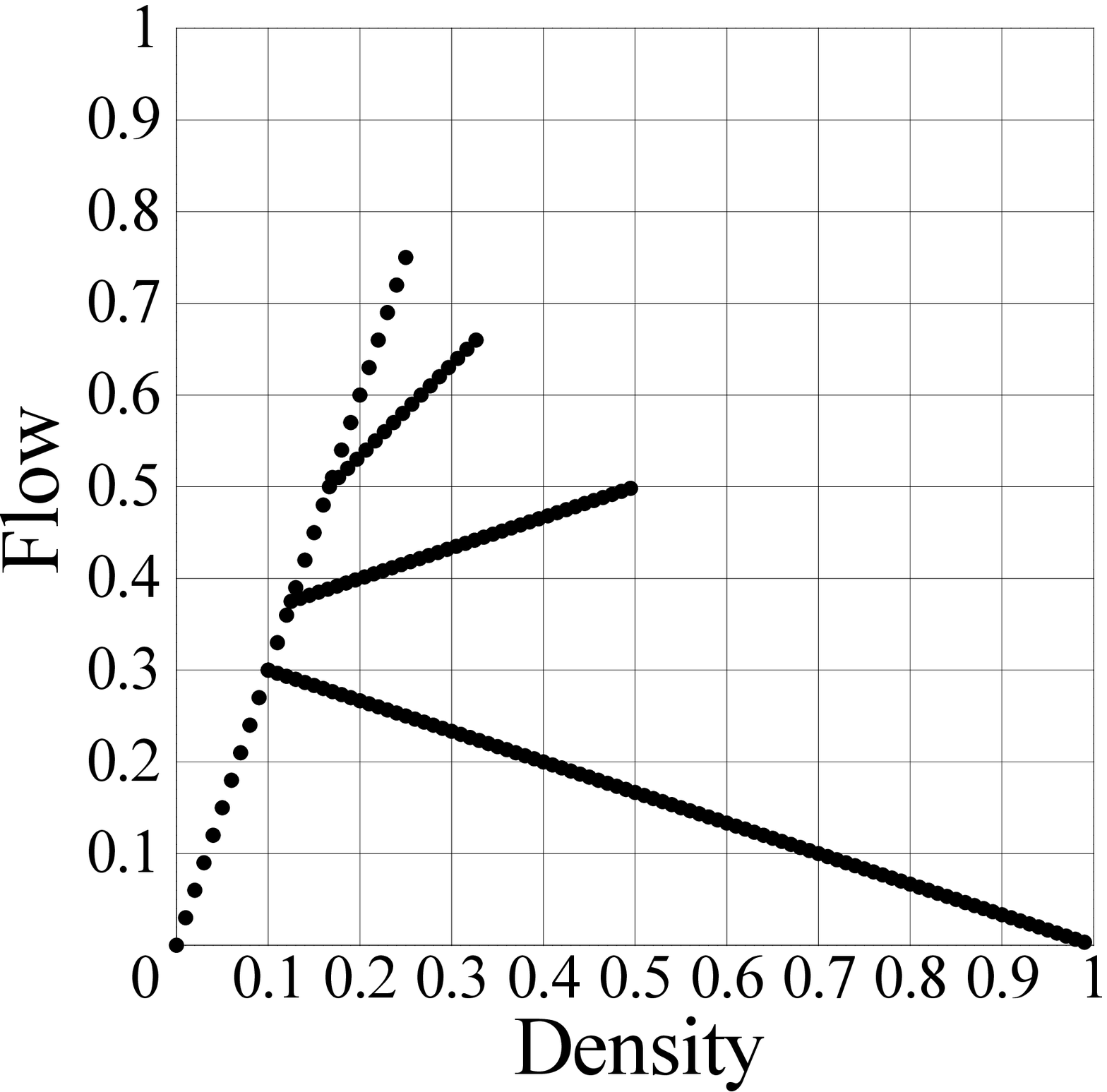}
\end{center}
\caption{The spatio-temporal pattern (left)
and the fundamental diagram (right) of the s2s--OVCA~\cite{Oguma2009}.
The number of the cells $L$ is fixed at $L=100$ 
and the periodic boundary condition is imposed.
The number of the cars in the above spatio-temporal pattern is $K=30$.
The maximum velocity $v_0$ and the monitoring
period $n_0$ are $v_0=3$ and $n_0=2$.
The flows $Q$ are computed by averaging over the time
period $800\leq n\leq 1000$, in which the traffic is expected to be in
equilibrium.}
\label{fig:1}
\end{figure}

Figure~\ref{fig:1} gives typical 
examples of the spatio-temporal pattern showing jams and the
fundamental diagram of the s2s--OVCA~\cite{Oguma2009}.
The spatio-temporal pattern shows the trajectories of
vehicles. Though irregular patterns are observed at the
initial stage of time evolution, trajectories become regular in the long run,
which consist of 
several jam clusters moving backward at the same velocity and 
vehicles running at the top speed.
The fundamental diagram, or the flow--density relation, in other words,  
gives the relation between the vehicle flow
\[
  Q:=\dfrac{1}{(n_f-n_i+1)L}\sum_{k=1}^K\sum_{n=n_i}^{n_f}
  \dfrac{x_k^{n+1}-x_k^n}{\delta t},
\]
which is equivalent to the total momentum of
vehicles per unit length averaged over the period $n_i\leq n\leq n_f$
when the traffic is expected to be in equilibrium,
and the vehicle density $\rho:=\frac{K}{Lx_0}$ where $L$ is 
the total number of the cells, or the length of the circuit,
on which the periodic boundary condition is imposed.
The fundamental diagram has the inverse-$\lambda$ shape with
several phases of traffic, namely, free, jam 
as well as metastable slow traffics,
which captures the characteristic
of observed flow--density relations~\cite{Chowdhury2000,Helbing2001}.
A comment here might be in order.
A CA-type model that is different from the s2s--OVCA
showing a fundamental diagram with several branches
was reported.~\cite{Nishinari2004} But all of its branches except for
the free line have negative inclinations. This feature is also different
from that of the fundamental diagram of the s2s--OVCA.
As we can
see in Fig.~\ref{fig:1}, the fundamental diagram consists of straight
branches, whose number, four in this case, is the same as that 
of all possible integral velocities, $v=0, 1, 2$ and $3(=v_0)$. 
The main result of this paper is to present 
a set of exact solutions of the s2s--OVCA, which
explains the piecewise linear fundamental diagram
of the s2s--OVCA. 

The outline is as follows.
In \S\ref{sec:1}, the fundamental diagram given in Fig.~\ref{fig:1} 
will be examined. An empirical formula for the fundamental diagram
of the s2s--OVCA will be introduced. A set of exact solutions of the
s2s--OVCA that reproduces the empirical formula will be presented
in \S\ref{sec:2} and \S\ref{sec:3}. Concluding remarks are
given in the final section. 
Derivation of the s2s--OVCA~\eref{eq:s2s--OVCA} from the s2s--OV 
model~\eref{eq:s2s--OV} is briefly summarized in an appendix.

\section{Numerical Observation of the Fundamental Diagram}
\label{sec:1}
\begin{figure}[h]
\label{fig:2}
\begin{center}
\includegraphics[width=60mm]{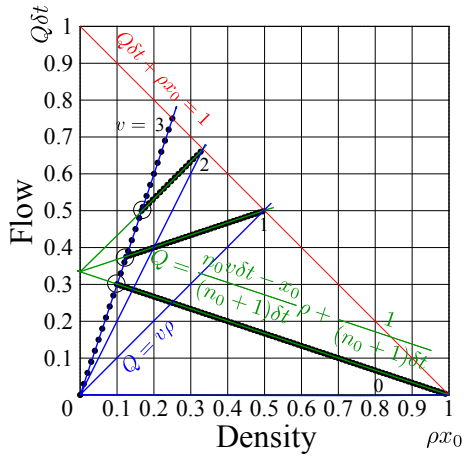}
\end{center}
\caption{(Color online) An empirical formula for the fundamental diagram of the
s2s--OVCA. All the dots on the diagram were numerically
computed~\cite{Oguma2009} for the case
in which the maximum velocity $v_0$ and the monitoring
period $n_0$ are $v_0=3$ and $n_0=2$.
Each solid line corresponds to
the formula in the same color or in the same grayscale.
Each branch is labeled with a number that shows the
integral velocity $v$ corresponding to the branch.
The end points of the branches are determined by the maximum
density for each integral velocity. The branch points of the slow
or the jamming lines, which are encircled with thin small circles, 
give the lowest densities for these lines.}
\end{figure}

By numerical experiments~\cite{Oguma2009}, the
fundamental diagram of the s2s--OVCA was observed to be
piecewise linear and possesses $v_0$ branches irrespective of
the parameter $n_0$. Let us have a close look at the
fundamental diagram in Figs.~\ref{fig:1} and \ref{fig:2} and 
examine these features more in detail.

As shown in Fig.~\ref{fig:2},
each branch corresponds to an integral
velocity that is less than or equal to the maximum velocity
$v_0$.
All the end points of the branches are observed to be 
on the line
\begin{equation}
  Q\delta t+\rho x_0=1.
  \label{eq:max_density} 
\end{equation}
This is because
the density at the end point is determined by
the maximum density $\rho_{\rm max}(v)$
that allows the velocity of the slowest 
car to be $v$. Because of the time evolution of 
the s2s--OVCA~\eref{eq:s2s--OVCA}, a car running at a velocity $v$
must have an interval that is larger than or equal to $v\delta t+x_0$.
Imagining a traffic such that all the cars flow 
at the same velocity $v$ with the minimum interval $v \delta t+x_0$,
we conclude that
the maximum density $\rho_{\rm max}(v)$ is determined by
\begin{equation}
  \rho_{\rm max}(v)=\dfrac{1}{v\delta t+x_0},
  \label{eq:max_density1}
\end{equation}
since each car occupies $v\delta t+x_0$ cells.
The flow corresponding to the traffic is given by
\begin{equation}
  Q\bigl(\rho_{\rm max}(v)\bigr)=v\rho_{\rm max}(v).
  \label{eq:free_line1}
\end{equation}
Equations~\eref{eq:max_density1} and \eref{eq:free_line1} simultaneously
holds on the end points and they yield 
$Q\bigl(\rho_{\rm max}(v)\bigr)\delta t+\rho_{\rm max}(v)x_0=1$. Thus 
all the end points must satisfy the relation~\eref{eq:max_density}.

The free line is a branch that agrees with the
straight line whose 
inclination equals to the maximum velocity $v_0$,
\begin{equation}
  Q=v_0\rho,
  \label{eq:free_line}
\end{equation}
since all the cars on the free line flow at the maximum speed
$v_0$.
Any other branches branch off from the free line. By observation,
the density of the branching point of the branch corresponding
to a velocity $v$ reads
\begin{equation}
  \rho_{\rm min}(v)=\dfrac{1}{n_0(v_0-v)\delta t+v_0\delta t +x_0}.
  \label{eq:min_density}
\end{equation}
This observation is explained as follows. Suppose 
one car, say the car $k$, 
runs at the velocity $v$ and all the other $K-1$ cars
run at the maximum velocity $v_0$.
At the moment the $k$-th car
slows down to $v$, the interval between
the cars $k$ and $k+1$ is $v\delta t +x_0$. Since
it takes at least $n_0+1$ time steps for the car $k$ to
speed up to $v_0$, the interval between the cars $k$ and $k+1$
expands up to $H=(n_0+1)(v_0-v)\delta t+v\delta t+x_0
=1/\rho_{\rm min}(v)> v_0\delta t$ by the time the $k$-th
car speeds up to $v_0$. If all the cars can obtain the
interval $H$, slow cars running at the velocity $v$ 
should vanish in the end. Thus the density at 
the branching point of the branch corresponding
to the velocity $v\delta t$
is given by $\rho_{\rm min}(v)=1/H$, 
which is nothing but the minimum
density of the branch.

Suppose the slow branches form straight lines. Then these lines
must be the segments connecting the branching point
$\bigl(\rho_{\rm min}(v), v_0\rho_{\rm min}(v)\bigr)$ 
and the end point 
$\bigl(\rho_{\rm max}(v), v\rho_{\rm max}(v)\bigr)$.
The expression for these line segments is obtained
by eqs.~\eref{eq:max_density1}, 
\eref{eq:free_line1} 
and \eref{eq:min_density} as
\begin{equation}
  Q = \dfrac{n_0 v\delta t-x_0}{(n_0+1)\delta t}\rho
  +\dfrac{1}{(n_0+1)\delta t},
  \label{eq:empirical}
\end{equation}
which agrees fairly well with 
the branches for the slow traffics 
consisting of the numerically obtained dots in Fig.~\ref{fig:2}.
Coincidence of the $Q$-interception of slow branches
$\frac{1}{(n_0+1)\delta t}$,
which also appears in Fig.~\ref{fig:2},
is also shown in the above empirical formula.

The aim of the following two sections is to explain
the empirical formula for the fundamental diagram~\eref{eq:empirical}
based on exact solutions of the s2s--OVCA.

\section{Solutions with a Single Slow Cluster}
\label{sec:2}

After transient flow goes on evolving irregularly for a while,
the traffic reaches ``equilibrium'' in the long run where 
slow-traffic clusters of seemingly invariant sizes move regularly,
as one can see in the spatio-temporal pattern in Fig.~\ref{fig:1}.
Here we allow equilibrium to change sizes of slow or fast clusters
periodically but prohibit it from increasing or 
decreasing their sizes monotonically.
We shall investigate such equilibrium flows and their flow--density
relations. Limiting ourselves to simple cases, we begin with 
equilibrium flows with, at most, a single slow cluster.

To begin with, we shall see a specific spatio-temporal pattern
starting from the following initial configuration,
\begin{equation}
  0:~\verb*|1 2 3 4       5    |.
  \label{eq:example}
\end{equation}
Note that the number 0 at the leftmost shows the time.
The digits and the blank symbols \verb*| | in the above configuration
mean the indices of the cars and the empty cells, respectively.
We set the monitoring period $n_0$ at 2.
The velocity of the cars 4 and 5 is 3, which is the top velocity $v_0$ 
of this case. The velocity of the cars 1, 2 and 3, whose headways are 1, is 1.
Then in this case, 
the headways of all the cars in the past have nothing to do with
the time evolution starting from the above configuration~\eref{eq:example}. 
The periodic boundary condition is assumed so 
the length of the circuit is 19 in this case.

Using eq.~\eref{eq:s2s--OVCA2}, which is equivalent to
the s2s--OVCA~\eref{eq:s2s--OVCA} with $x_0=\delta t=1$,
we can confirm that the following spatio-temporal pattern evolves
out of the above initial configuration~\eref{eq:example},
\begin{eqnarray}
  &0:& \verb*|1 2 3 4       5    | \nonumber \\
  &1:& \verb*| 1 2 3   4       5 | \label{eq:example2} \\
  &2:& \verb*|5 1 2 3     4      | \nonumber \\
  &3:& \verb*| 5 1 2 3       4   | \nonumber.
\end{eqnarray}
Moving all the cells of the initial configuration
one cell rightward as well as neglecting the difference of the car indices,
we get the configuration at the time 3. In this sense, the
initial configuration~\eref{eq:example} gives a periodic motion of the vehicles
and hence an equilibrium flow of the s2s--OVCA~\eref{eq:s2s--OVCA}.
Roughly speaking, we shall construct equilibrium flows of the above type
in the following.

The most simple equilibrium flow is the free traffic in which
all the cars always run at the top velocity $v_0$.
In other words, the free traffic is a traffic flow
with no slow cluster. The headway
of a car running at the top velocity must be larger than or equal to
$v_0\delta t$ and so is each headway in the free traffic.
Since all the cars share the same velocity in the free traffic,
all the headways are invariant. Thus the free traffic is in
equilibrium. One concludes that
the flow--density relation for the free traffic is given 
by the free line~\eref{eq:free_line} because of its definition
given above.

A snapshot of an equilibrium flow with only a single slow 
cluster is given by a configuration of vehicles shown below,
\begin{equation}
  \underbrace{
  |\stackrel{1}{\langle v\rangle}\stackrel{[0]}{\rightarrow}\cdots
  \stackrel{k-1}{\langle v\rangle}
  \stackrel{[0]}{\rightarrow}
  \stackrel{k}{\langle v\rangle}
  \stackrel{[0]}{\leadsto}
  \stackrel{k+1}{\langle v_0\rangle}
  \stackrel{[n_0+1]}{\rightarrow}\cdots
  \stackrel{K-1}{\langle v_0\rangle}
  \stackrel{[n_0+1]}{\rightarrow}
  \stackrel{K}{\langle v_0\rangle}
  \stackrel{[m+1]-lx_0}{\leadsto}}_{\displaystyle L}
  |\stackrel{1}{\langle v\rangle},
  \label{eq:equilibrium_initial}
\end{equation}
where $m=0,1,\cdots,n_0$ and $l x_0/\delta t=0,1\cdots,v_0-v-1$.
The car $n$, $n=1,2,\cdots, K$, which will move 
at a velocity $v$, $v=0,1,\cdots,v_0$, in the next turn,
is denoted by $\stackrel{n}{\langle v\rangle}$.
Thus the cars 1 to $k$ form a single
slow cluster in the above configuration and the cars
$k+1$ to $K$ form a single fast cluster. Thus the integer $k=0,1,
\cdots, K$ gives the number of cars in the slow cluster.
We shall not deal with the case $k=0$ since its corresponding
configuration gives a free traffic without
slow clusters. Thus $k$ is limited to $k=1,2,\cdots,K$ in the following.

An arrow $\stackrel{h}{\rightarrow}$, $h=0,1,2,\cdots$, 
is the headway whose number is $h$, or $h$ empty cells
between the neighboring cars. 
The symbol $[n]$ that appears on the arrows
means $[n]=v\delta t+n(v_0-v)\delta t$.
Two headways in front of the cars $k$ and $K$,
which are the tops of the slow and the fast clusters respectively,
are denoted by wavy arrows $\leadsto$ in order to stress
that they are going to change by $(v_0-v)\delta t$ at each step.
Dots between a pair of cars
with the same velocity $v$ and headway $[h]$, 
$\stackrel{n_1}{\langle v\rangle}\stackrel{[h]}{\rightarrow}\cdots
\stackrel{n_2}{\langle v\rangle}\stackrel{[h]}{\rightarrow}$, 
mean that all the cars and the headways in between also have the
same velocity and headway, 
$\stackrel{n}{\langle v\rangle}\stackrel{[h]}{\rightarrow}$, 
$n_1\leq {}^\forall n \leq n_2$.
The vertical bars mark the initial cell of the car 1. 
The car 1 and the vertical bar, which appear again on the rightmost,
show the periodicity of the circuit and they will be omitted 
in the following. 
Note that the initial configuration~\eref{eq:example}
is given by the snapshot~\eref{eq:equilibrium_initial} with
$v_0=3,\ v=1,\ n_0=2,\ K=5,\ k=3$ and $l=3$,
\[
  0:~\verb*|1 2 3 4       5    |=
  |\stackrel{1}{\langle 1\rangle}\stackrel{1}{\rightarrow}
  \stackrel{2}{\langle 1\rangle}\stackrel{1}{\rightarrow}
  \stackrel{3}{\langle 1\rangle}\stackrel{1}{\leadsto}
  \stackrel{4}{\langle 3\rangle}\stackrel{7}{\rightarrow}
  \stackrel{5}{\langle 3\rangle}\stackrel{4}{\leadsto},
\]
where $1=[0]$, $7=[3]$, $4=[3]-3$ and $x_0=\delta t=1$.

The circuit length $L$, which is shown under the brace
in the above configuration, is then given by
\begin{equation}
  L=L_0-(n_0-m)(v_0-v)\delta t-lx_0,
  \label{eq:circuit_length}
\end{equation}
where $L_0:=(K-k)(n_0+1)(v_0-v)\delta t
+K\bigl(x_0+v\delta t\bigr)$.
With use of $\rho_{\rm min}(v)$ given in eq.~\eref{eq:min_density},
the formula for the circuit length~\eref{eq:circuit_length} is 
cast into 
\begin{equation}
  k(n_0+1)(v_0-v)\delta t=\dfrac{K}{\rho_{\rm min}(v)}-L
  -(n_0-m)(v_0-v)\delta t-l x_0,
  \label{eq:circuit_length2}
\end{equation}
where $1/\rho_{\rm min}(v)$ is an integer.
Since the parameter $k$ gives the number of cars in a slow cluster,
it must be non-negative. 
Equation \eref{eq:circuit_length2} gives the
condition for $k$ to be non-negative as
\[
  0\leq\dfrac{K}{\rho_{\rm min}(v)}-L
  -(n_0-m)(v_0-v)\delta t-l x_0.
\]
Thus we obtain the necessary condition for $k$ to be non-negative as
\begin{equation}
  \rho:=\dfrac{K}{L}\geq\rho_{\rm min}(v),
  \label{eq:min_density2}
\end{equation}
since $\frac{(n_0-m)(v_0-v)\delta t+l x_0}{L}\geq 0$.
The lower bound of $\rho$ in the above inequality
coincides with the numerically obtained 
minimum density~\eref{eq:min_density} for the branch
corresponding to a velocity $v$.

We should confirm here that there always exist non-negative
integers $k, l$ and $m$ satisfying eq.~\eref{eq:circuit_length}
for arbitrarily given $L, K, n_0, v_0$ and $v$ that meet
the condition~\eref{eq:min_density2}.
The r.h.s.~of eq.~\eref{eq:circuit_length2} must be divided by
$(n_0+1)(v_0-v)\delta t$. Since the third and fourth terms
cover all the residue classes modulo $(n_0+1)(v_0-v)\delta t$,
there always exist some non-negative integers $k, l$ and $m$ 
satisfying the above equation for 
arbitrary $L, K, n_0, v_0$ and $v$.

The time evolution out of the 
initial configuration \eref{eq:equilibrium_initial} gives an
equilibrium flow with a single slow cluster. 
Let us confirm it by
checking each step of the time evolution. We exclude the case
$k=K$ and shall deal with it separately.
Watching the initial state again, 
\begin{equation}
  0:
  |\stackrel{1}{\langle v\rangle}\stackrel{[0]}{\rightarrow}\cdots
  \stackrel{k-1}{\langle v\rangle}
  \stackrel{[0]}{\rightarrow}
  \stackrel{k}{\langle v\rangle}
  \stackrel{[0]}{\leadsto}
  \stackrel{k+1}{\langle v_0\rangle}
  \stackrel{[n_0+1]}{\rightarrow}\cdots
  \stackrel{K-1}{\langle v_0\rangle}
  \stackrel{[n_0+1]}{\rightarrow}
  \stackrel{K}{\langle v_0\rangle}
  \stackrel{[m+1]-lx_0}{\leadsto},
  \label{eq:equilibrium0}
\end{equation}
where $0$ at the leftmost shows the time, 
we notice that the
cars 1 to $k$ run at a slow velocity $v$ and the cars $k+1$ to $K$
run at the top velocity $v_0$. Thus only two
headways denoted by wavy arrows for clarity, 
namely those of the cars $k$ and $K$ that are 
the tops of the slow cluster and the fast cluster respectively, 
are going to change by $(v_0-v)\delta t$ at each step.
The configuration of the next time step is then given by
\[
  1:
  |\stackrel{v\delta t}{\Rightarrow}
  \stackrel{1}{\langle v\rangle}\stackrel{[0]}{\rightarrow}\cdots
  \stackrel{k-1}{\langle v\rangle}
  \stackrel{[0]}{\rightarrow}
  \stackrel{k}{\langle v\rangle}_1
  \stackrel{[1]}{\leadsto}
  \stackrel{k+1}{\langle v_0\rangle}
  \stackrel{[n_0+1]}{\rightarrow}\cdots
  \stackrel{K-1}{\langle v_0\rangle}
  \stackrel{[n_0+1]}{\rightarrow}
  \stackrel{K}{\langle v_0\rangle}
  \stackrel{[m]-lx_0}{\leadsto}.
\]
The subscript 1 of the car $k$ shows
that it is the first turn for the car $k$ to have a headway 
which is long enough, $[1]=v_0\delta t$,
to run at the top speed $v_0$. 
Similarly, a subscript $n$ of the $k$-th car, 
$\stackrel{k}{\langle v\rangle}_n$, $v<v_0$, 
whose headway is longer than or equal to $v_0\delta t$, 
means that it is the $n$-th turn for the car to have 
its headway long enough to run at
the maximum velocity $v_0$. 
This symbol is going to appear shortly and to be
used in the same meaning in the following.
The thick arrow at the leftmost
shows the displacement of the car 1 from its initial cell.
Until the car $K$ decelerates at the $m$-th time step,
the cars 1 to $k$ run at a slow velocity $v$ and all the others
run at the top speed $v_0$.
Thus the configuration at the $n$-th
time step, $1\leq n \leq m-1$, is expressed by
\begin{equation}
  n: 
  |\stackrel{nv\delta t}{\Rightarrow}
  \stackrel{1}{\langle v\rangle}\stackrel{[0]}{\rightarrow}\cdots
  \stackrel{k-1}{\langle v\rangle}
  \stackrel{[0]}{\rightarrow}
  \stackrel{k}{\langle v\rangle}_n
  \stackrel{[n]}{\leadsto}
  \stackrel{k+1}{\langle v_0\rangle}
  \stackrel{[n_0+1]}{\rightarrow}\cdots
  \stackrel{K-1}{\langle v_0\rangle}
  \stackrel{[n_0+1]}{\rightarrow}
  \stackrel{K}{\langle v_0\rangle}
  \stackrel{[m+1-n]-lx_0}{\leadsto}.
  \label{eq:equilibrium1}
\end{equation}
At the $m$-th time step, the headway of the car $K$
shortens to $[1]-lx_0=v_0\delta t-lx_0$ and its velocity
immediately slows down to $v_0-lx_0/\delta t$. Thus
there additionally appears another varying headway 
denoted by a wavy arrow in front of the car $K-1$.
The configuration at the $m$-th time step is then given by
\begin{equation}
  m:
  |\stackrel{m v\delta t}{\Rightarrow}
  \stackrel{1}{\langle v\rangle}\stackrel{[0]}{\rightarrow}\cdots
  \stackrel{k-1}{\langle v\rangle}
  \stackrel{[0]}{\rightarrow}
  {\stackrel{k}{\langle v\rangle}}_{m}\hspace*{-1em}
  \stackrel{[m]}{\leadsto}
  \stackrel{k+1}{\langle v_0\rangle}
  \stackrel{[n_0+1]}{\rightarrow}\cdots
  \stackrel{K-1}{\langle v_0\rangle}
  \stackrel{[n_0+1]}{\leadsto}
  \stackrel{K}{\langle v_0-lx_0/\delta t\rangle}
  \stackrel{[1]-lx_0}{\leadsto}.
  \label{eq:equilibrium3}
\end{equation}
At this moment, velocities of the cars $K-1$, $K$ and 1 are
respectively $v_0$, $v_0-lx_0/\delta t$ and $v$, as well as 
the headways of the cars $K-1$ and $K$ irregularly shortens by $lx_0$
and $(v_0-v)\delta t-lx_0$. 
We classify the car $K$ with the intermediate velocity
$v_0-lx_0/\delta t$ into the fast cluster.
The slow cluster with the velocity $v$ consists of cars with
only a single slow velocity $v$.
Thus the car $K$ has been the top
of the fast cluster by this turn. But in the next turn, 
the car $K$ also slows down to $v$
and becomes the bottom of the slow cluster. Thus the
configuration at the $(m+1)$-th time step is
\[
  m+1:
  |\stackrel{(m+1) v\delta t}{\Rightarrow}
  \stackrel{1}{\langle v\rangle}\stackrel{[0]}{\rightarrow}\cdots
  \stackrel{k-1}{\langle v\rangle}
  \stackrel{[0]}{\rightarrow}
  {\stackrel{k}{\langle v\rangle}}_{m+1}\hspace*{-2em}
  \stackrel{[m+1]}{\leadsto}
  \stackrel{k+1}{\langle v_0\rangle}
  \stackrel{[n_0+1]}{\rightarrow}\cdots
  \stackrel{K-2}{\langle v_0\rangle}
  \stackrel{[n_0+1]}{\rightarrow}
  \stackrel{K-1}{\langle v_0\rangle}
  \stackrel{[n_0+1]-lx_0}{\leadsto}
  \stackrel{K}{\langle v\rangle}
  \stackrel{[0]}{\rightarrow}.
\]
Until the car $k$ accelerates at the $(n_0+1)$-th time step,
$k+1$ cars, namely
the cars 1 to $k$ as well as the car $K$, 
run at a slow velocity $v$ and all the others
run at the top speed $v_0$.
Thus the configuration at the $n$-th
time step, $m+1\leq n\leq n_0$, is expressed by
\begin{equation}
  n:
  |\stackrel{n v\delta t}{\Rightarrow}
  \stackrel{1}{\langle v\rangle}\stackrel{[0]}{\rightarrow}\cdots
  \stackrel{k-1}{\langle v\rangle}
  \stackrel{[0]}{\rightarrow}
  {\stackrel{k}{\langle v\rangle}}_{n}
  \stackrel{[n]}{\leadsto}
  \stackrel{k+1}{\langle v_0\rangle}
  \stackrel{[n_0+1]}{\rightarrow}\cdots
  \stackrel{K-2}{\langle v_0\rangle}
  \stackrel{[n_0+1]}{\rightarrow}
  \stackrel{K-1}{\langle v_0\rangle}
  \stackrel{[n_0-n+m+2]-lx_0}{\leadsto}
  \stackrel{K}{\langle v\rangle}
  \stackrel{[0]}{\rightarrow}.
  \label{eq:equilibrium2}
\end{equation}
At the $(n_0+1)$-th time step, the car $k$ accelerates to
the top velocity $v_0$ since 
its headway has kept no less than $v_0\delta t$ for $n_0$
time steps by the time.
Thus the car $k$ switches to the fast cluster and the
cars 1 to $k-1$ as well as the car $K$ now form the slow cluster. 
The place of the varying headway denoted by a wavy arrow also 
moves from the car $k$ to the car $k-1$. Thus the configuration
at the $(n_0+1)$-th time step looks as
\begin{equation}
  n_0+1:
  |\stackrel{(n_0+1) v\delta t}{\Rightarrow}
  \stackrel{1}{\langle v\rangle}\stackrel{[0]}{\rightarrow}\cdots
  \stackrel{k-1}{\langle v\rangle}
  \stackrel{[0]}{\leadsto}
  {\stackrel{k}{\langle v_0\rangle}}
  \stackrel{[n_0+1]}{\rightarrow}
  \stackrel{k+1}{\langle v_0\rangle}
  \stackrel{[n_0+1]}{\rightarrow}\cdots
  \stackrel{K-2}{\langle v_0\rangle}
  \stackrel{[n_0+1]}{\rightarrow}
  \stackrel{K-1}{\langle v_0\rangle}
  \stackrel{[m+1]-lx_0}{\leadsto}
  \stackrel{K}{\langle v\rangle}
  \stackrel{[0]}{\rightarrow}.
  \label{eq:equilibrium6}
\end{equation}
Since the vehicles are running on the circuit of the length $L$,
we may put the car $K$ behind the car 1. Then the 
configuration at the $(n_0+1)$-th time step
\begin{equation}
  n_0+1:
  |\stackrel{n_0v\delta t-x_0}{\Rightarrow}
  \stackrel{K}{\langle v\rangle}
  \stackrel{[0]}{\rightarrow}
  \stackrel{1}{\langle v\rangle}\stackrel{[0]}{\rightarrow}\cdots
  \stackrel{k-1}{\langle v\rangle}
  \stackrel{[0]}{\leadsto}
  {\stackrel{k}{\langle v_0\rangle}}
  \stackrel{[n_0+1]}{\rightarrow}\cdots
  \stackrel{K-2}{\langle v_0\rangle}
  \stackrel{[n_0+1]}{\rightarrow}
  \stackrel{K-1}{\langle v_0\rangle}
  \stackrel{[m+1]-lx_0}{\leadsto},
  \label{eq:equilibrium5}
\end{equation}
becomes the same as the initial configuration up to 
the rightward displacement of $n_0v\delta t-x_0$ cells.
Thus the motion of vehicles
evolving out of the initial configuration~\eref{eq:equilibrium0} is
periodic with the period $n_0+1$, which shows the corresponding
traffic is in equilibrium.

The configuration for the case $k=K$
\[
  n:|\stackrel{nv\delta t}{\Rightarrow}
  \stackrel{1}{\langle v\rangle}
  \stackrel{[0]}{\rightarrow}
  \cdots
  \stackrel{K}{\langle v\rangle}
  \stackrel{[0]}{\rightarrow}
\]
gives a constant traffic flow in which all the cars run at the
slow velocity $v$ with the same headway $[0]=v\delta t$.
Thus we confirm that the case $k=K$ also 
gives an equilibrium traffic.

We should note that the fast cluster in the case $k=K-1$ 
temporally vanishes from the $(m+1)$-th time step
to the $n_0$-th time step,
\begin{equation}
\begin{split}
  0:|& 
  \stackrel{1}{\langle v\rangle}\stackrel{[0]}{\rightarrow}\cdots
  \stackrel{K-2}{\langle v\rangle}
  \stackrel{[0]}{\rightarrow}
  \stackrel{K-1}{\langle v\rangle}
  \stackrel{[0]}{\leadsto}
  \stackrel{K}{\langle v_0\rangle}
  \stackrel{[m+1]-lx_0}{\leadsto},\\
  n: 
  |&\stackrel{nv\delta t}{\Rightarrow}
  \stackrel{1}{\langle v\rangle}\stackrel{[0]}{\rightarrow}\cdots
  \stackrel{K-2}{\langle v\rangle}
  \stackrel{[0]}{\rightarrow}
  \stackrel{K-1}{\langle v\rangle}_n
  \stackrel{[n]}{\leadsto}
  \stackrel{K}{\langle v_0\rangle}
  \stackrel{[m+1-n]-lx_0}{\leadsto},
  \quad 1\leq n\leq m-1, \\
  m: 
  |&\stackrel{mv\delta t}{\Rightarrow}
  \stackrel{1}{\langle v\rangle}\stackrel{[0]}{\rightarrow}\cdots
  \stackrel{K-2}{\langle v\rangle}
  \stackrel{[0]}{\rightarrow}
  \stackrel{K-1}{\langle v\rangle}_m
  \stackrel{[m]}{\leadsto}
  \stackrel{K}{\langle v_0-lx_0/\delta t\rangle}
  \stackrel{[1]-lx_0}{\leadsto},\\
  n:
  |&\stackrel{n v\delta t}{\Rightarrow}
  \stackrel{1}{\langle v\rangle}\stackrel{[0]}{\rightarrow}\cdots
  \stackrel{K-2}{\langle v\rangle}
  \stackrel{[0]}{\rightarrow}
  \stackrel{K-1}{\langle v\rangle}_n
  \stackrel{[m+1]-lx_0}{\leadsto}
  \stackrel{K}{\langle v\rangle}
  \stackrel{[0]}{\rightarrow},
  \quad m+1\leq n\leq n_0.
\end{split}
\label{eq:equilibrium7}
\end{equation}
But it appears appears again at the $(n_0+1)$-th time step
\[
  n_0+1:|\stackrel{(n_0+1)v\delta t}{\Rightarrow}
  \stackrel{1}{\langle v\rangle}\stackrel{[0]}{\rightarrow}\cdots
  \stackrel{K-3}{\langle v\rangle}
  \stackrel{[0]}{\rightarrow}
  \stackrel{K-2}{\langle v\rangle}
  \stackrel{[0]}{\leadsto}
  \stackrel{K-1}{\langle v_0\rangle}
  \stackrel{[m+1]-lx_0}{\leadsto}
  \stackrel{K}{\langle v\rangle}
  \stackrel{[0]}{\leadsto},
\]
which goes back to the shifted initial configuration,
\[
  n_0+1:|\stackrel{n_0v\delta t-x_0}{\Rightarrow}
  \stackrel{K}{\langle v\rangle}
  \stackrel{[0]}{\rightarrow}
  \stackrel{1}{\langle v\rangle}
  \stackrel{[0]}{\rightarrow}
  \cdots
  \stackrel{K-3}{\langle v\rangle}
  \stackrel{[0]}{\rightarrow}
  \stackrel{K-2}{\langle v\rangle}
  \stackrel{[0]}{\leadsto}
  \stackrel{K-1}{\langle v_0\rangle}
  \stackrel{[m+1]-lx_0}{\leadsto}.
\]
Thus such vanishment of the fast cluster as we have observed above
does not break equilibrium of the traffic.

Now let us compute the flow of this periodic motion averaged over the period
$n_0+1$.
As we have already seen in the above time evolutions for the
cases $k=1,2,\cdots, K-1$, $k$ and $K-k$ cars respectively run 
at the velocities $v$ and $v_0$ in $0\leq n\leq m-1$. 
At the $m$-th time step, 
$k$ and $K-k-1$ cars run at the velocities $v$ and $v_0$ and remaining
one car move at $v_0-lx_0/\delta t$. In the following period,
$m+1\leq n\leq n_0$, $k+1$ and $K-k-1$ cars move 
at the velocities $v$ and $v_0$, respectively. 
Thus the average flow $Q$ is 
\begin{align*}
  Q & = \dfrac{1}{(n_0+1)L}\Bigl(
  m\bigl(kv+(K-k)v_0\bigr) +\bigl(kv+(K-k-1)v_0+v_0-lx_0/\delta t\bigr)\\
  & \quad + (n_0-m)\bigl((k+1)v+(K-k-1)v_0\bigr)  \Bigr)\\
  & = \dfrac{n_0 v\delta t-x_0}{(n_0+1)\delta t}\rho
  +\dfrac{1}{(n_0+1)\delta t},
\end{align*}
which is the same as the empirical formula for 
the flow--density relation~\eref{eq:empirical}.
For the case $k=K$, the flow $Q$ and the density $\rho$ are
respectively given by
\[
  Q=\rho v, \quad \rho=\dfrac{1}{v\delta t+x_0}.
\]
A straightforward calculation shown below
\begin{align*}
  Q & = \rho\dfrac{(n_0+1)v\delta t}{(n_0+1)\delta t}
    = \dfrac{(n_0v\delta t-x_0)+(v\delta t+x_0)}{(n_0+1)\delta t}\rho \\
    & = \dfrac{n_0v\delta t-x_0}{(n_0+1)\delta t}\rho
    + \dfrac{1}{(n_0+1)\delta t}
\end{align*}
proves that the above two
quantities are related by the flow--density relation~\eref{eq:empirical}.

The above formula is interpreted in a different and more intuitive manner.
The configurations~\eref{eq:equilibrium0} and \eref{eq:equilibrium5}
shows the rightward displacement of
the entire configuretion~\eref{eq:equilibrium0} 
by $n_0v\delta t-x_0$ cells in $n_0+1$ 
time steps. The flow provided by this motion of the entire
configuration gives the first 
term $\frac{n_0 v\delta t-x_0}{(n_0+1)\delta t}\rho$
of the above formula. Here we should remind ourselves of
the leftward displacement of the car $K$ by $L$ cells, namely
whole the circuit length, which is fictitiously introduced
to make the shifted initial configuration~\eref{eq:equilibrium5}
from the real configuration~\eref{eq:equilibrium6} at the $(n_0+1)$-th
discrete time. In order to compensate the underestimation of the
flow caused by this fictitious displacement, 
we have to add the flow corresponding to
the rightward displacement of the car $K$ by $L$ cells
in $n_0+1$ time steps
\[
  \dfrac{1}{(n_0+1)L}\cdot\dfrac{L}{\delta t}
  =\dfrac{1}{(n_0+1)\delta t},
\]
which agrees with the second term of the above formula for the
average flow.

\section{Solutions with Multiple Slow Clusters}
\label{sec:3}

As we have observed in the spatio-temporal pattern in Fig.~\ref{fig:1}, 
several slow or jam clusters that share the same slow velocity coexist
in equilibrium. Here we make solutions as such. 

Roughly speaking, we can make an equilibrium solution with 
multiple slow clusters by putting several snapshots of the
equilibrium solutions in the previous section together.
Let us see how this idea works via observation of a specific
example. Thanks to the periodic boundary condition, we can
move the leftmost empty cell of 
the configuration~\eref{eq:example2} to the rightmost,
\begin{eqnarray*}
  \verb*| 1 2 3   4       5 | & = & 
  |\stackrel{1}{\Rightarrow}\verb*|1 2 3   4       5  | \\
  & = &|\stackrel{1}{\Rightarrow}
  \stackrel{1}{\langle 1\rangle}\stackrel{1}{\rightarrow}
  \stackrel{2}{\langle 1\rangle}\stackrel{1}{\rightarrow}
  \stackrel{3}{\langle 1\rangle}_1\stackrel{3}{\leadsto}
  \stackrel{4}{\langle 3\rangle}\stackrel{7}{\rightarrow}
  \stackrel{5}{\langle 2\rangle}\stackrel{2}{\leadsto}.
\end{eqnarray*}
Putting the above configuration and 
the configuration~\eref{eq:example}
\[
  \verb*|1 2 3 4       5    |=|
  \stackrel{1}{\langle 1\rangle}\stackrel{1}{\rightarrow}
  \stackrel{2}{\langle 1\rangle}\stackrel{1}{\rightarrow}
  \stackrel{3}{\langle 1\rangle}\stackrel{1}{\leadsto}
  \stackrel{4}{\langle 3\rangle}\stackrel{7}{\rightarrow}
  \stackrel{5}{\langle 3\rangle}\stackrel{4}{\leadsto}
\]
together, we get a configuration with two slow clusters as
\begin{eqnarray}
  &&\verb*|1 2 3   4       5  6 7 8 9       0    |\nonumber\\
  &&\qquad =
  \stackrel{1}{\langle 1\rangle}\stackrel{1}{\rightarrow}
  \stackrel{2}{\langle 1\rangle}\stackrel{1}{\rightarrow}
  \stackrel{3}{\langle 1\rangle}_1\stackrel{3}{\leadsto}
  \stackrel{4}{\langle 3\rangle}\stackrel{7}{\rightarrow}
  \stackrel{5}{\langle 2\rangle}\stackrel{2}{\leadsto}
  \stackrel{6}{\langle 1\rangle}\stackrel{1}{\rightarrow}
  \stackrel{7}{\langle 1\rangle}\stackrel{1}{\rightarrow}
  \stackrel{8}{\langle 1\rangle}\stackrel{1}{\leadsto}
  \stackrel{9}{\langle 3\rangle}\stackrel{7}{\rightarrow}
  \stackrel{0}{\langle 3\rangle}\stackrel{4}{\leadsto}
  \label{eq:example3}
\end{eqnarray}
Using eq.~\eref{eq:s2s--OVCA2}, we can confirm that
the time evolution out of the above configuration gives
a spatio-temporal patterns below,
\begin{eqnarray*}
  &0:& \verb*|1 2 3   4       5  6 7 8 9       0    | \\
  &1:& \verb*| 1 2 3     4      5 6 7 8   9       0 | \\
  &2:& \verb*|0 1 2 3       4    5 6 7 8     9      | \\
  &3:& \verb*| 0 1 2   3       4  5 6 7 8       9   |.
\end{eqnarray*}
Moving all the cells of the initial configuration
one cell rightward as well as neglecting the difference of the car 
indices, we get the configuration at the time 3. Thus the
configuration~\eref{eq:example3} gives a periodic motion of the vehicles
and hence an equilibrium flow of the s2s--OVCA~\eref{eq:s2s--OVCA}.
In the following, we shall confirm that the above construction
of equilibrium flows with multiple slow clusters works
in general, too.

As building blocks of such
solutions, we introduce symbols denoting slow and fast clusters in the
snapshots of the equilibrium solutions \eref{eq:equilibrium0}, 
\eref{eq:equilibrium1}, \eref{eq:equilibrium3} and \eref{eq:equilibrium2}
in the previous section,
\begin{equation}
  n:|\stackrel{d}{\Rightarrow}
  \langle S\rangle
  \stackrel{h_{s}}{\leadsto}
  \langle F \rangle\stackrel{h_{f}}{\leadsto},\quad
  d=\left\{\begin{array}{ll}
    nv\delta t & 0\leq n\leq m \\
    (n-1)v\delta t-x_0 & m+1\leq n\leq n_0
  \end{array}\right.,
  \label{eq:slowandfastclusters}
\end{equation}
where $\langle S\rangle\stackrel{h_{s}}{\leadsto}$ and 
$\langle F\rangle\stackrel{h_{f}}{\leadsto}$
respectively denote the slow and the fast clusters 
with varying headways $\stackrel{h_{s,f}}{\leadsto}$ whose lengths
are $h_{s,f}$ on their tops.
For instance, the slow and fast clusters in the 
snapshot of the $m$-th time step~\eref{eq:equilibrium3} are
respectively given by
\begin{align*}
  &  \langle S\rangle\stackrel{h_{s}}{\leadsto} := 
  \stackrel{1}{\langle v\rangle}\stackrel{[0]}{\rightarrow}\cdots
  \stackrel{k-1}{\langle v\rangle}
  \stackrel{[0]}{\rightarrow}
  {\stackrel{k}{\langle v\rangle}}_{m}\hspace*{-1em}
  \stackrel{[m]}{\leadsto} \\
  &  \langle F\rangle_F\stackrel{h_{f}}{\leadsto} := 
  \stackrel{k+1}{\langle v_0\rangle}
  \stackrel{[n_0+1]}{\rightarrow}\cdots
  \stackrel{K-1}{\langle v_0\rangle}
  \stackrel{[n_0+1]}{\leadsto}
  \stackrel{K}{\langle v_0-lx_0/\delta t\rangle}
  \stackrel{[1]-lx_0}{\leadsto},
\end{align*}
where $h_s=[m]$ and $h_f=[1]-lx_0$.
Note that the car $K$ with the intermediate velocity
$v_0-lx_0/\delta t$ is classified into the fast cluster.
We also note that we can always rearrange the order of the vehicles
into the form of eq.~\eref{eq:slowandfastclusters}, thanks to
the periodic boundary condition.

An equilibrium solution with $N$ slow clusters are made with the
slow and fast clusters~\eref{eq:slowandfastclusters} with the same
monitoring period $n_0$, the slow velocity $v$ and the top
velocity $v_0$. First, we prepare $N$ pairs of slow and fast clusters in the
configurations of $N$ equilibrium solutions with only a single 
slow cluster, 
$\langle S_i \rangle\stackrel{h_{s_i}}{\leadsto}$, 
$\langle F_i \rangle\stackrel{h_{f_i}}{\leadsto}$
$i=1,2,\cdots,N$. 
We shall consider a configuration made by putting these pairs of slow and fast 
clusters in line,
\begin{equation}
  \langle S_{1} \rangle\stackrel{h_{s_1}}{\leadsto}
  \langle F_{1} \rangle\stackrel{h_{f_1}}{\leadsto}
  \langle S_{2} \rangle\stackrel{h_{s_2}}{\leadsto}
  \langle F_{2} \rangle\stackrel{h_{f_2}}{\leadsto}
  \cdots
  \langle S_{N} \rangle\stackrel{h_{s_N}}{\leadsto}
  \langle F_{N} \rangle\stackrel{h_{f_N}}{\leadsto}.
  \label{eq:multi_initial}
\end{equation}
where the periodic boundary condition is imposed.
Thus the total length
of the circuit is the sum of the lengths of all the clusters.

We observed that vanishment of a fast cluster~\eref{eq:equilibrium7}
does not break equilibrium because
the vanished fast cluster was revived by acceleration of
the top car of the slow cluster behind.
However, the top car of the fast cluster does not slow down
if the slow cluster in front of it vanishes.
Thus a vanishing slow cluster does break equilibrium.
If slow down to the slow velocity $v$ of
the top car of any fast cluster $\langle F_i\rangle$ is no later than
speed up to the top velocity $v_0$ of the bottom car of the slow
cluster $\langle S_{i+1}\rangle$ in front of 
the fast cluster $\langle F_i\rangle$, 
such vanishment of slow clusters is prevented. Thus we require
the above ``no vanishing slow cluster'' (NVSC) rule to be fulfilled
by any pairs of the fast and slow clusters, 
$\langle F_i\rangle$ and $\langle S_{i+1}\rangle$, in the
configurations \eref{eq:multi_initial}.

We shall see below that
the above configuration~\eref{eq:multi_initial} satisfying
the NVSC rule
gives an equilibrium solution with $N$ slow clusters.
Speaking more specifically,
we shall confirm that the motion of vehicles belonging to a
pair of slow and fast clusters $\langle S_{i} \rangle
\stackrel{h_{s_{i}}}{\leadsto}\langle F_{i} 
\rangle\stackrel{h_{f_{i}}}{\leadsto}$, $i=1,2,\cdots,N$, in
the  above configuration~\eref{eq:multi_initial}
with multiple slow clusters is the same as that without 
any other pairs of slow and fast clusters
we have investigated in detail in the previous section.

Consider the time evolution of a fast cluster 
and its varying headway
$\langle F_{i} \rangle\stackrel{h_{f_{i}}}{\leadsto}$, $i=1,2,\cdots,N$.
It is determined
by the motion of the vehicle at the bottom of the slow cluster
$\langle S_{i+1} \rangle\stackrel{h_{s_{i+1}}}{\leadsto}$
in front of it,
which is different from its original partner in the solution
with a single slow-cluster
$\langle S_{i}\rangle\stackrel{h_{s_{i}}}{\leadsto}$ behind it.
However, thanks to the NVSC rule, the bottom of 
$\langle S_{i+1} \rangle\stackrel{h_{s_{i+1}}}{\leadsto}$
keeps the slow velocity $v$ until the top of
$\langle F_{i} \rangle\stackrel{h_{f_{i}}}{\leadsto}$
slows down to $v$. Thus the motion of the bottom of
$\langle S_{i+1} \rangle\stackrel{h_{s_{i+1}}}{\leadsto}$
is the same as that of the bottom of
$\langle S_{i}\rangle\stackrel{h_{s_{i}}}{\leadsto}$ and
so is the motion of the fast cluster
$\langle F_{i} \rangle\stackrel{h_{f_{i}}}{\leadsto}$
until its top slows down to $v$.
Let us now think about the time evolution
after the top of
$\langle F_{i} \rangle\stackrel{h_{f_{i}}}{\leadsto}$
slows down to $v$. In the same manner we observed in the solution
with only a single slow cluster in the previous section, 
the top of $\langle F_{i} \rangle\stackrel{h_{f_{i}}}{\leadsto}$
keeps the slow velocity $v$ at least for
$n_0+1$ time steps.
The time evolution of the
varying headway, which locates in front of the second car of 
$\langle F_{i} \rangle\stackrel{h_{f_{i}}}{\leadsto}$, 
and the motion of the remaining cars of 
$\langle F_{i} \rangle\stackrel{h_{f_{i}}}{\leadsto}$ are now
determined by the above motion of the top of 
$\langle F_{i} \rangle\stackrel{h_{f_{i}}}{\leadsto}$.
Thus the time evolution of 
the fast cluster and its varying headway
$\langle F_{i} \rangle\stackrel{h_{f_{i}}}{\leadsto}$ remains 
the same as that with only a single slow cluster we observed
in the previous section at least for
$n_0+1$ time steps after the top of
$\langle F_{i} \rangle\stackrel{h_{f_{i}}}{\leadsto}$
slows down to $v$ and at least for $n_0+1$ time steps
from the beginning.

Let us turn ourselves now to 
the time evolution of a slow cluster and its varying headway 
$\langle S_{i} \rangle
\stackrel{h_{s_{i}}}{\leadsto}$, $i=1,2,\cdots,N$.
The time evolution is determined
by the motion of the vehicle at the bottom of the fast cluster 
in front of it,
which is nothing but its original partner in the solution
with a single slow-cluster, $\langle F_{i} 
\rangle\stackrel{h_{f_{i}}}{\leadsto}$. We now know that
the time evolution of the fast cluster $\langle F_{i} 
\rangle\stackrel{h_{f_{i}}}{\leadsto}$ is the
same as those with only a single slow cluster for $n_0+1$ time steps.
Thus the time evolution of the the slow cluster 
$\langle S_{i} \rangle\stackrel{h_{s_{i}}}{\leadsto}$
and its varying headway in the 
configuration
with multiple slow clusters~\eref{eq:multi_initial} is the same as that 
in the configuration with only a single slow cluster
for $n_0+1$ time steps.

As we saw in the configuration with only a single slow cluster
in the previous section, the time evolution of
the slow and the fast clusters as well as their varying headways are
periodic with the period $n_0+1$. Thus,
also in the configuration
with multiple slow clusters~\eref{eq:multi_initial},
the time evolution of 
all of the slow and the fast clusters as well as their varying
headways are periodic with the period $n_0+1$. 
To summarize, we have confirmed that 
the motion of vehicles belonging to a
pair of slow and fast clusters $\langle S_{i} \rangle
\stackrel{h_{s_{i}}}{\leadsto}\langle F_{i} 
\rangle\stackrel{h_{f_{i}}}{\leadsto}$, $i=1,2,\cdots,N$, in
the configuration~\eref{eq:multi_initial}
with multiple slow clusters is the same as that without 
any other pairs of slow and fast clusters
we have investigated in detail in the previous section and that
it gives an equilibrium solution with $N$ slow clusters.

Let us examine the flow--density relation of the traffic evolving out
of the configuration with $N$ slow clusters~\eref{eq:multi_initial}.
Let $L_i$, $K_i$, $Q_i$ and $\rho_i:=K_i/L_i$ be the circuit length, the number
of vehicles,
the mean flow averaged over one period, i.e.
$n_0+1$ time steps, and the density of vehicles for the configuration
with only one slow cluster, 
$\langle S_i\rangle\stackrel{h_{s_{i}}}{\leadsto}\langle F_i\rangle\stackrel{h_{f_{i}}}{\leadsto}$, 
$i=1,2,\cdots,N$. In \S\ref{sec:2}, we confirmed that $Q_i$ and $\rho_i$
satisfied the flow--density relation~\eref{eq:empirical},
\begin{equation}
  Q_i=\dfrac{n_0 v\delta t-x_0}{(n_0+1)\delta t}\rho_i
  +\dfrac{1}{(n_0+1)\delta t}.
  \label{eq:empirical2}
\end{equation}
The circuit length $L$ and the number of vehicles $K$ 
of the configuration with
$N$ slow clusters are then expressed by
$L:=\sum_{i=1}^N L_i$ and $K:=\sum_{i=1}^N K_i$.
The mean flow $Q$ that is again averaged over the period
as well as the vehicle density $\rho$
of the configuration with $N$ slow clusters are expressed by
\begin{align*}
  Q & = \dfrac{1}{L}\sum_{i=1}^N L_i Q_i=\sum_{i=1}^{N}\lambda_i Q_i, \\
  \rho & = \dfrac{K}{L} =\sum_{i=1}^N \lambda_i\rho_i,
\end{align*}
with $\lambda_i:=L_i/L$. Using an identity $\sum_{i=1}^N\lambda_i=1$
as well as the flow density relation for the configuration with
a single slow cluster~\eref{eq:empirical2}, one 
straightforwardly confirms
\begin{align*}
  Q-\dfrac{n_0 v\delta t-x_0}{(n_0+1)\delta t}\rho & =
  \sum_{i=1}^N\lambda_i\Bigl(
      Q_i-\dfrac{n_0 v\delta t-x_0}{(n_0+1)\delta t}\rho_i
  \Bigr) \\
  & = \Bigl(\sum_{i=1}^N\lambda_i\Bigr)\dfrac{1}{(n_0+1)\delta t} 
  = \dfrac{1}{(n_0+1)\delta t},
\end{align*}
which concludes that
$Q$ and $\rho$ also satisfy the flow--density relation~\eref{eq:empirical}.
Note that the relation can be confirmed in a more intuitive manner, 
as we have shown for the configuration with only a single slow
cluster.

\section{Concluding Remarks}
\label{sec:4}
We have introduced a set of exact solutions of the s2s--OVCA and
have explained the piecewise linear fundamental diagram of the model.
The empirical formula for the flow--density relation of 
the s2s--OVCA~\eref{eq:empirical} has been
read out of the numerically obtained fundamental diagram in Figs.~\ref{fig:1}
and \ref{fig:2}. A set of exact periodic solutions that possesses only
a single slow cluster has been presented, which has reproduced the 
formula for the flow--density relation of the s2s--OVCA~\eref{eq:empirical}.
With the use of these solutions, another set of exact periodic solutions
with multiple slow clusters, which have been observed in the numerically
obtained spatio-temporal pattern in Fig.~\ref{fig:1} has been
introduced. And these solutions again have explained the flow--density
relation of the s2s--OVCA~\eref{eq:empirical}.

We should comment about a couple of problems in our mind 
related to the present results.
The solutions given in \S\ref{sec:2} and \S\ref{sec:3} provide
a set of equilibrium flows which correspond all the points
on the fundamental diagram of the s2s--OVCA. But we do not know
if the set we have presented in \S\ref{sec:2} and \S\ref{sec:3}
gives all possible equilibrium flows or not.
There might be a solution giving an equilibrium flow that does not belong to
the set. By numerical experiments~\cite{Oguma2009}, we observed
that time evolutions out of all the initial conditions we 
verified eventually go to some equilibrium flows. But this observation
still remains numerical observation and lacks proof.
For the case of $n_0=1$, 
these problems were investigated~\cite{Tian2009} and
should be extended to all the cases of $n_0$'s.
As was shown in our previous paper~\cite{Oguma2009},
the s2s--OVCA~\eref{eq:s2s--OVCA} transforms to
the s2s--OV model~\eref{eq:s2s--OV} through
the inverse ultradiscretization and the continuous limit.
Note that the monitoring period $n_0$ in the
s2s--OVCA must be a free parameter so as to make
the monitoring period $t_0$ in the 
s2s--OV model a free parameter.
It should be clarified if the characteristics of the equilibrium
solution of the s2s--OVCA given in this paper remain in
the solution of the s2s--OV model. 

\acknowledgements One of the authors (HU) is grateful to 
K.~Oguma for the previous collaboration.

\appendix
\section{}
We shall derive the s2s--OVCA~\eref{eq:s2s--OVCA} from the s2s--OV 
model~\eref{eq:s2s--OV} in a brief way. 
Introducing the OV function
\[
  v_{\rm opt}(x):=v_0\Bigl(
  \dfrac{1}{1+\e^{-(x-x_0)/\delta x}}
  -\dfrac{1}{1+\e^{x_0/\delta x}}\Bigr)
\]
as well as the effective distance
\[
  \Delta_{\rm eff} x_k(t):=\delta x\log\Bigl(\dfrac{1}{t_0}\int_0^{t_0}
  \e^{-\Delta x_k(t-t^\prime)/\delta x}\d t^\prime\Bigr)^{-1},
\]
the s2s--OV model is expressed as
\begin{equation}
  \dfrac{\d x_k(t)}{\d t} 
  = v_{\rm opt}\bigl(\Delta_{\rm eff} x_k(t)\bigr).
  \label{eq:s2s-OV_app}
\end{equation}
Let $x_k^n:=x_k(t=n\delta t)$ and $v_k^n:=(x_k^{n+1}-x_k^n)/\delta t$.
Discretization of the effective distance $\Delta_{\rm eff} x_k(t)$ is then
expressed as
\[
  \Delta_{\rm eff}^{\rm d}x_k^n:=\delta x\log\Bigl(\sum_{n^\prime=0}^{n_0}
  \dfrac{\e^{-\Delta x_k^{n-n^\prime}/\delta x}}{n_0+1}\Bigl)^{-1},
\]
where $n_0:=t_0/\delta t$. Using the fact that the OV function 
$v_{\rm opt}(x)$ is written as a limit of a function
\[
  v_{\rm opt}^{\rm d}(x):=
  \dfrac{\delta x}{\delta t}\log
    \biggl[
      \dfrac{1+\e^{(x-x_0)/\delta x}}{1+\e^{-x_0/\delta x}}
      \bigg/\dfrac{1+\e^{(x-x_0-v_0\delta t)/\delta x}}
      {1+\e^{-(x_0+v_0\delta t)/\delta x}}
    \biggr]
\]
as $v_{\rm opt}(x)=\lim_{\delta t\rightarrow 0}v_{\rm opt}^{\rm d}(x)$,
the s2s--OV model~\eref{eq:s2s-OV_app} in a 
time-discretized form is given by
\begin{equation}  
  v_k^n=v_{\rm opt}^{\rm d}\bigl(\Delta_{\rm eff}^{\rm d} x_k^{n}\bigr),
  \label{eq:ds2s-OV}
\end{equation}
which is equivalent to
\begin{align*}
  x_k^{n+1}& =x_k^n+\delta x\Biggl\{
  \log\biggl[
      1+\Bigl(\sum_{n^\prime=0}^{n_0} 
      \dfrac{\e^{-(\Delta x_k^{n-n^\prime}-x_0)/\delta x}}{n_0+1}
      \Bigr)^{-1}\biggr]
  -\log\bigl(1+\e^{-x_0/\delta x}\bigr) \\
  & \qquad\qquad -
  \log\biggl[
      1+\Bigl(\sum_{n^\prime=0}^{n_0} 
      \dfrac{\e^{-(\Delta x_k^{n-n^\prime}-x_0-v_0\delta t)/\delta x}}{n_0+1}
      \Bigr)^{-1}\biggr]
  +\log\bigl(1+\e^{-(x_0+v_0\delta t)/\delta x}\bigr)
  \biggr\}.
\end{align*}
It is straightforward to confirm
that the continuum limit $\delta t\rightarrow 0$ of the above
discrete s2s--OV model~\eref{eq:ds2s-OV} reduces 
to eq.~\eref{eq:s2s-OV_app} or equivalently the s2s--OV 
model \eref{eq:s2s--OV}.

Ultradiscretization~\cite{Tokihiro1996} is a scheme 
for getting a piecewise-linear equation from 
a difference equation via the limit formula
\[
  \lim_{\delta x\rightarrow +0}\delta x
  \log(\e^{A/\delta x}+\e^{B/\delta x}+\cdots)
  =\max(A,B,\cdots).
\]
In the ultradiscrete limit, the OV function for 
the discrete s2s--OV model $v_{\rm opt}^{\rm d}(x)$ goes to
\begin{equation*}
  \lim_{\delta x\rightarrow  +0}v_{\rm opt}^{\rm d}(x)
  =:v_{\rm opt}^{\rm u}(x)
  = \max\Bigl(0,\dfrac{x-x_0}{\delta t}\Bigr)
  -\max\Bigl(0,\dfrac{x-x_0}{\delta t} -v_0\Bigr),
\end{equation*}
which is nothing but the OV function for the uOV model~\cite{Takahashi2009}.
The effective distance $\Delta_{\rm eff}^{\rm d}x_k^n$ 
on the other hand is also ultradiscretized in the same manner:
\begin{equation*}
    \Delta_{\rm eff}^{\rm u}x_k^n:=\lim_{\delta x\rightarrow +0}
    \Delta_{\rm eff}^{\rm d}x_k^n
    = -\max_{n^\prime=0}^{n_0}\bigl(-\Delta x_k^{n-n^\prime}\bigr) 
    = \min_{n^\prime=0}^{n_0}\bigl(\Delta x_k^{n-n^\prime}\bigr).
\end{equation*}
Thus we obtain an ultradiscrete equation
\begin{equation}
  v_k^n=v_{\rm opt}^{\rm u}\bigl(\Delta_{\rm eff}^{\rm u} x_k^{n}\bigr),
  \label{eq:us2s--OV}
\end{equation}
which is equivalent to
\begin{equation*}
  x_k^{n+1}= x_k^n+\max\Bigl(0,
  \min_{n^\prime=0}^{n_0}\bigl(\Delta x_k^{n-n^\prime}\bigr)-x_0\Bigr) 
  -\max\Bigl(0,
  \min_{n^\prime=0}^{n_0}\bigl(\Delta x_k^{n-n^\prime}\bigr)-x_0-v_0\delta t
  \Bigr),
\end{equation*}
as the ultradiscrete limit of the discrete s2s--OV model~\eref{eq:ds2s-OV}.
We name it the ultradiscrete s2s--OV model.

Now let us see how the s2s--OVCA comes out from the ultradiscrete
s2s--OV model~\eref{eq:us2s--OV}.
Assume that the headway between 
the cars $k$ and $k+1$, $\widetilde{\Delta}x_k^n:=\Delta x_k^n -x_0$,
for any $k$ must be non-negative, $\widetilde{\Delta}x_k^n\geq 0$,
which prohibits car-crash from now on.
Then the ultradiscrete s2s--OV model~\eref{eq:us2s--OV} reduces to 
the s2s--OVCA~\eref{eq:s2s--OVCA}.
Note that the constants $x_0$ and $v_0\delta t$ can be arbitrary
in the above derivation. We set $x_0=1$ and $v_0\delta t$ at an integer
so that the positions of the vehicles $x_k^n$ become integers at any time.


\begin{thebibliography}{99}
\bibitem{Chowdhury2000} D.~Chowdhury, L.~Santen and A.~Schadschneider:
Phys.~Rep.~{\bf 329} (2000) 199.

\bibitem{Helbing2001} D.~Helbing:
Rev.~Mod.~Phys. {\bf 73} (2001) 1067.

\bibitem{Bando1995} M.~Bando, K.~Hasebe, A.~Nakayama, A.~Shibata 
and Y.~Sugiyama:
Phys.~Rev.~E {\bf 51} (1995) 1035.

\bibitem{Nagel1992} K.~Nagel and M.~Schreckenberg:
J.~Physique I {\bf 2} (1992) 2221.

\bibitem{Wolfram1986} S.~Wolfram:
{\it Theory and Applications of Cellular Automata}
(World Scientific, Singapore, 1986).

\bibitem{Fukui1996} M.~Fukui and Y.~Ishibashi:
J.~Phys.~Soc.~Jpn.~{\bf 65} (1996) 1868.

\bibitem{Takayasu1993} M.~Takayasu and H.~Takayasu:
Fractals {\bf 1} (1993) 860.

\bibitem{Takahashi2009} D.~Takahashi and J.~Matsukidaira:
JSIAM Letters {\bf 1} (2009) 1.

\bibitem{Tokihiro1996} T.~Tokihiro, D.~Takahashi,
J.~Matsukidaira and J.~Satsuma:
Phys.~Rev.~Lett.~{\bf 76} (1996) 3247.

\bibitem{Kanai2009} M.~Kanai, S.~Isojima, K.~Nishinari and 
T.~Tokihiro: Phys.~Rev.~E {\bf 79} (2009) 056108.

\bibitem{Takahashi1997} D.~Takahashi and J.~Matsukidaira:
J.~Phys.~A: Math.~Gen.~{\bf 30} (1997) L733.

\bibitem{Emmerich1998} H.~Emmerich, T.~Nagatani and K.~Nakanishi:
Physica A~{\bf 254} (1998) 548.

\bibitem{Oguma2009} K.~Oguma and H.~Ujino:
JSIAM Letters {\bf 1} (2009) 68.

\bibitem{Newell1961} G.~F.~Newell:
Oper.~Res.~{\bf 9} (1961) 209.

\bibitem{Nishinari2004} K.~Nishinari, M.~Fukui and A.~Schadschneider:
J.~Phys.~A.~Math.~Gen.~{\bf 37} (2004) 3101.

\bibitem{Tian2009} R.~Tian:
Disc.~Appl.~Math.~{\bf 157} (2009) 2904.
\end{thebibliography}
\end{document}